\documentclass[sigconf,natbib=true,anonymous=false]{acmart}

\AtBeginDocument{%
  }

\usepackage{hyperref}

\setcopyright{acmlicensed}
\copyrightyear{2025}
\acmYear{2025}
\acmDOI{10.1145/3705328.3759328}

\acmConference[RecSys '25]{the 19th ACM Recommender Systems Conference}{September 22--26,
  2025}{Prague, Czech Republic}

\begin{document}

\title{Leveraging Geometric Insights in Hyperbolic Triplet Loss for Improved Recommendations}

\author{Viacheslav Yusupov}
\affiliation{%
  \institution{HSE University}
  \city{Moscow}
  \country{Russian Federation}
}
\email{vyusupov@hse.ru}

\author{Maxim Rakhuba}
\affiliation{%
  \institution{HSE University}
  \city{Moscow}
  \country{Russian Federation}}

\author{Evgeny Frolov}
\affiliation{%
  \institution{AIRI}
  \city{Moscow}
  \country{Russian Federation}
}
\affiliation{%
  \institution{HSE University}
  \city{Moscow}
  \country{Russian Federation}
}

\begin{teaserfigure}
    \centering
    \includegraphics[width=0.71\linewidth]{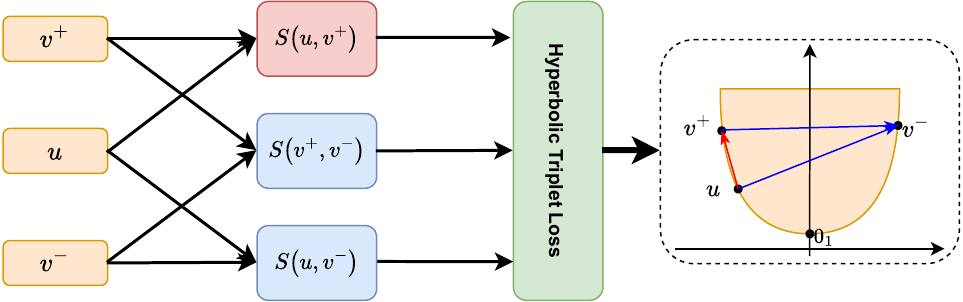}
    \caption{The Hyperbolic Triplet Loss function minimizes the hyperbolic distance from a user $u$ to the user's positive item $v^+$, while simultaneously maximizing the distance to the user's  negative item $v^-$, as well as maximizing the distance between the positive and negative items. This effect is visually demonstrated on a $2D$ hyperboloid at the top of the scheme with the corresponding vectors. The function $S(\cdot,\cdot)$ denotes the hyperbolic distance. In the rightmost part of the illustration, the red distance is minimized, while the blue distances are maximized.}
    \label{fig:model}
\end{teaserfigure}

\begin{abstract}
Recent studies have demonstrated the potential of hyperbolic geometry for capturing complex patterns from interaction data in recommender systems. In this work, we introduce a novel hyperbolic recommendation model that uses geometrical insights to improve representation learning and increase computational stability at the same time. We reformulate the notion of hyperbolic distances to unlock additional representation capacity over conventional Euclidean space and learn more expressive user and item representations. To better capture user-items interactions, we construct a triplet loss that models ternary relations between users and their corresponding preferred and nonpreferred choices through a mix of pairwise interaction terms driven by the geometry of data. Our hyperbolic approach not only outperforms existing Euclidean and hyperbolic models but also reduces popularity bias, leading to more diverse and personalized recommendations. 
\end{abstract}




\maketitle
\section{Introduction}

Recommender systems have a significant impact on various aspects of modern digital life, including e-commerce, social networks, online streaming platforms, and more \cite{alamdari2020systematic,anandhan2018social}.
To improve the quality of recommendations, modern models learn dependencies between users and items using elaborate representational learning techniques. These include methods such as \emph{BPR} \cite{rendle2012bpr}, metric learning approaches \cite{hsieh2017collaborative, liu2019user}, and neural network-based methods \cite{he2020lightgcn, kang2018self}. While the mentioned models successfully utilize embeddings in Euclidean space, recent research has demonstrated the effectiveness of representations in the hyperbolic space\cite{kruk}. The latter offers higher capacity and is better suited for non-trivially distributed data, such as user-item interactions \cite{peng2021hyperbolic}. For example, the model \cite{chamberlain2019scalable} incorporates hyperbolic \emph{BPR} \cite{rendle2012bpr}. \cite{frolov2024self,mirvakhabova2020performance} presents a hyperbolic modification of \emph{SASRec} \cite{kang2018self}. 

Although many hyperbolic recommendation models primarily learn representations through user-item interactions \cite{guo2023hyperbolic}, there is a limited amount of works that also incorporate item-item interactions within hyperbolic space. Inspired by the success of item-based interaction learning using Euclidean embeddings \cite{xue2019deep}, we enhance our hyperbolic approach by integrating an item-to-item component.

In this work, we introduce an effective \emph{TriplH} hyperbolic approach with Lorentzian triplet loss that incorporates ternary item-user-item interactions leveraging geometric insights (see Figure \ref{fig:model}). Our model learns more comprehensive user and item vector representations compared to conventional Euclidean embeddings, resulting in a significant boost in performance metrics, as demonstrated in our experiments. Therefore, to achieve the target quality of the recommendations, a significantly smaller embedding size is required in the hyperbolic space compared to the Euclidean space, which confirms previous insights \cite{kruk}. This approach reduces the number of model parameters while improving computational efficiency, making the model better suited for deployment in high-traffic services \cite{peng2021hyperbolic}.

Unlike the fixed margin used in \emph{HyperML} \cite{vinh2020hyperml}, our item-item interaction term plays the role of a user-adaptive margin, allowing it to better capture user preferences. The item-item interaction term not only learns relationships between positive and negative examples for each user but also captures popularity-based dependencies among items within a recommender system's catalogue. As a result, our \emph{TriplH} approach mitigates popularity bias and offers more diverse recommendations. Additionally, by utilizing the Lorentz hyperbolic model \cite{zhu2023lorentzian} instead of the Poincaré ball \cite{nickel2017poincare}, our method achieves faster inference and greater computational stability due to a reduction in complex operations, touching the boundaries of  hardware arithmetic precision \cite{peng2021hyperbolic}.

Overall our contributions can be summarized as follows.
\begin{itemize}
    \item We propose a \emph{TriplH} hyperbolic model with geometrically inspired triplet loss to create more expressive user and item low-dimensional representations, therefore boosting the quality of recommendations.
    \item We develop an adaptive margin approach to mitigate the popularity bias and increase the diversity of recommendations.
    \item We utilize the Lorentz model operations to provide faster inference with more stable computations. 
\end{itemize}
The rest of the paper is organized as follows: Section \ref{rel} presents a brief review of related work, Section \ref{pre} provides the necessary preliminaries for our method, and Section \ref{met} details the \emph{TriplH} approach. The experimental setup, results, and conclusions are discussed in Sections \ref{set}, \ref{res}, and \ref{con}, respectively. The model implementation is open-sourced and is available online\footnote{Code: \url{https://github.com/YusupovV-Lab/TriplH}}.

\section{Related Work}
\label{rel}
Recent studies have demonstrated the effectiveness of hyperbolic geometry in representing hierarchical data, such as graphs \cite{liu2019hyperbolic} or recommender systems \cite{peng2021hyperbolic,kruk}. Some hyperbolic recommender systems \cite{vinh2020hyperml,frolov2024self,li2022hyperbolic} employ the more interpretable Poincaré ball model \cite{nickel2017poincare}, while others \cite{xu2020learning,zhu2023lorentzian,cheng2025large} utilize the Lorentz model \cite{nickel2018learning}, which offers greater stability. These hyperbolic models typically adapt traditional approaches \cite{rendle2012bpr,kang2018self,lightgcn} by incorporating hyperbolic embeddings to improve model expressiveness. For instance, \emph{HyperML} \cite{vinh2020hyperml} is a hyperbolic variant of \emph{Collaborative metric learning} \cite{hsieh2017collaborative} with a trainable margin \cite{li2020symmetric}. Nevertheless, \cite{frolov2024self,mirvakhabova2020performance} present hyperbolic modifications of the modern \emph{SASRec} \cite{kang2018self} and \cite{sun2021hgcf} uses \emph{NGCF} \cite{wang2019neural} architecture in hyperbolic space.

However, there are relatively few hyperbolic methods that explicitly model ternary item-user-item interactions \cite{yusupov2025knowledge, zhao2022hyperbolic}, similarly to Knowledge Graph Completion \cite{balazevic2019tucker}, which limits their ability to fully harness the potential of representation learning. In this work, we propose an effective model that ensures stable computations within the Lorentz space \cite{nickel2018learning}. Our approach leverages hyperbolic ternary interactions between users and items and introduces an adaptive item-item margin to improve recommendation performance.

\section{Preliminaries}
\label{pre}
In this section, we briefly review the main concepts of hyperbolic geometry and introduce essential background information.
\subsection{Hyperbolic Geometry}
Several isomorphic and interchangeable models of hyperbolic geometry exist \cite{zhao2022hyperbolic}. The Poincaré ball \cite{nickel2017poincare} is the standard model used more frequently, while the Lorentz model \cite{nickel2018learning} is considered more computationally stable \cite{peng2021hyperbolic}. As we show further, the structural properties of the latter can be leveraged to avoid problematic computations with hyperbolic functions. 

Define an $n$-dimensional hyperboloid $\mathcal{H}^{n, \beta} \subset \mathbb{R}^{n+1}$: 
\begin{displaymath}
    \mathcal{H}^{n, \beta} = \{x \in \mathbb{R}^{n+1}: \|x\|_L^2 = -\beta\},
\end{displaymath}
with the  Lorentzian inner product for $u, v \in \mathcal{H}^{n, \beta}$:
\begin{equation}\label{inner}
\langle u, v \rangle_L = -u_0v_0 + \sum_{i = 1}^nu_iv_i,\quad x_0 = \sqrt{\beta + \sum_{i = 1}^n x_i^2},
\end{equation}
where $\beta$ is a space curvature, $\|x\|_L^2 = \langle x, x \rangle_L$ is the induced vector norm, $x_0 = \langle 0_\beta, x \rangle_L$ is the zeroth component of a vector in the hyperbolic space and $0_\beta = (\beta, 0, ..., 0) \in \mathbb{R}^{n+1}$ is the origin vector of the hyperboloid.

The associated geodesic distance between points $x, y \in \mathcal{H}^{n,\beta}$ is defined as:
\begin{equation}
d_L(x, y) = arccosh(-\langle x, y \rangle_L).
\label{geodesic}
\end{equation}
To circumvent the numerical instability of the geodesic distance computation \eqref{geodesic}, caused by the usage of unstable hyperbolic trigonometric functions
\cite{nickel2017poincare}, we replace it with an approximation based on the Squared Lorentz distance \cite{xu2020learning}:
\begin{equation}
    d_L^2(x, y) = 2\beta - 2\langle x, y \rangle_L.
    \label{square}
\end{equation}
This distance satisfies nearly all Euclidean distance axioms, except for the triangle inequality:
\begin{equation}
d(x, y) \le d(x, z) + d(y, z),
\label{ineq}
\end{equation}
for any $d(\cdot, \cdot)$ and points $x,y, z$ in the hyperbolic space.

\begin{figure}
    \centering
    \includegraphics[width=0.95\linewidth]{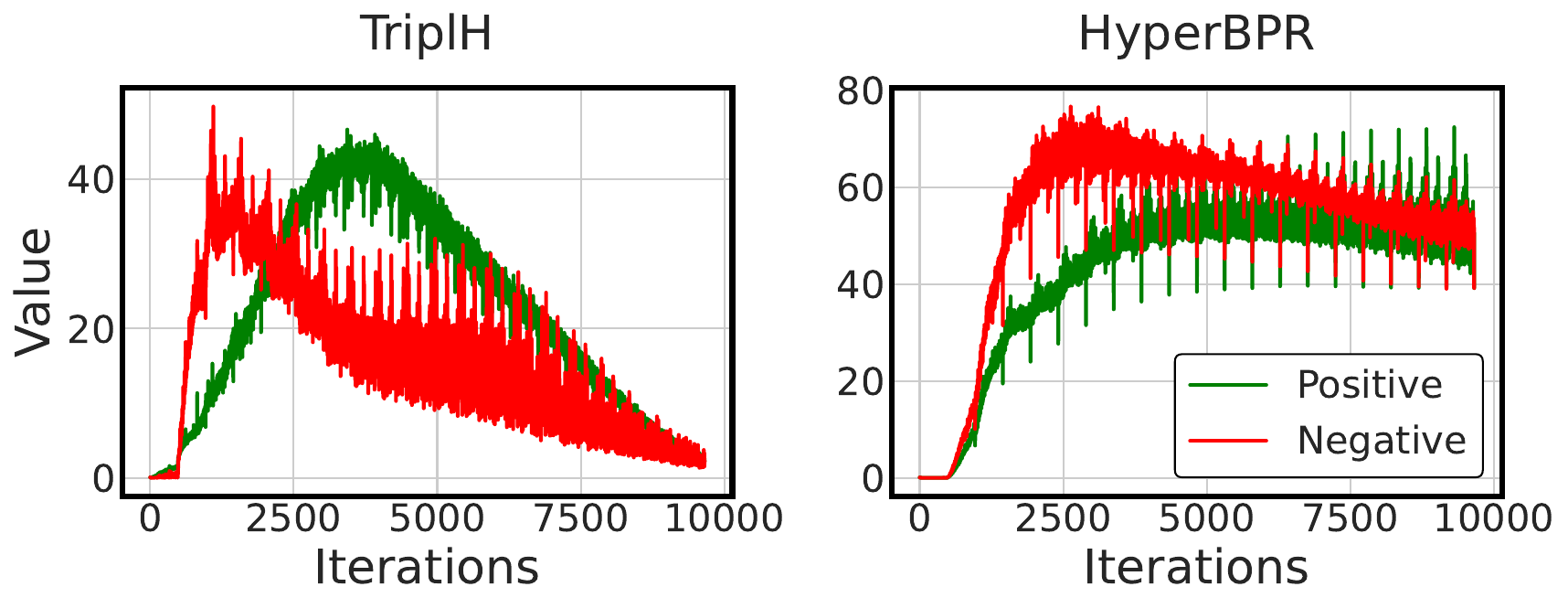}
    \vspace{-5pt}
    \caption{Distances from users to positive (green) and negative (red) items in \emph{TripleH} differ more compared to \emph{HyperBPR}.}
    \label{fig:comp}
\end{figure}

\subsection{Lorentzian Representations}
The representational advantages of the hyperbolic geometry for data with non-trivial distributional properties have been demonstrated in both graph representation learning \cite{liu2019hyperbolic} and recommender systems \cite{peng2021hyperbolic}. The \emph{LorentzFM} \cite{xu2020learning} approach focuses on learning pairwise relations via the sign of the triangle inequality \eqref{ineq} based on the Squared Lorentz distance \eqref{square}, rather than the hyperbolic distance itself. Formally, the \emph{LorentzFM} score function is defined as follows:
\begin{equation}
S(u, v) = \frac{1}{2} \cdot \frac{d^2_L(u, v) - d^2_L(0_1, u) - d_L^2(0_1, v)}{\langle 0_1, u \rangle_L \langle 0_1, v \rangle_L},
\label{lorentz}
\end{equation}
where $u \in \mathcal{H}^{d, \beta}$ and $v \in \mathcal{H}^{d, \beta}$ are user and item hyperbolic embeddings, respectively, with dimensionality $d+1$ and $0_{1} = (1, 0, ..., 0)$ is the origin vector for $\beta = 1$. A positive value of $S(u,v)$ indicates that points $u$ and $v$ are close to each other in the hyperbolic space, which suggests that item $v$ is suitable for the user $u$. 

The \emph{LorentzFM} model \cite{xu2020learning} is trained using Binary Cross-Entropy (BCE) loss with randomly sampled negative examples. However, recent studies have shown that Bayesian Personalized Ranking (BPR) loss \cite{rendle2012bpr} is also effective in hyperbolic space \cite{chamberlain2019scalable,vinh2020hyperml}.  Inspired by ideas from the \emph{LorentzFM} \cite{xu2020learning} approach, we also seek to learn user and item representations in the Lorentzian hyperbolic space, but \emph{extend it to the triplet-interaction setup utilizing geometric insights}. Our approach is presented in the following section.

\section{Proposed Approach}
\label{met}
In our work, we extend the concept of BPR loss in the hyperbolic space \cite{vinh2020hyperml} to a triplet loss that considers ternary interactions between the user, positive item, and negative item. Inspired by the effectiveness of ternary interactions in hyperbolic space \cite{yusupov2025knowledge}, we decompose the loss of our \emph{TriplH} model into pairwise interactions:
\begin{equation}
\begin{aligned}
    &L_{Triplet}(u, v^+, v^-) = \\
    &= \log \sigma\left(S(u, v^+) - S(u, v^-) - f(S(v^+, v^-))\right) + \lambda \langle v^+, v^-\rangle^2_L,
\end{aligned}
\label{loss}
\end{equation}
where $S(\cdot, \cdot)$ is a score function similar to \eqref{lorentz}, $u$ is a user embedding,  $v^+$ is a positive item embedding, and $v^-$ stands for a randomly sampled negative item. Similarly to  \cite{vinh2020hyperml}, we utilize $f(x) = ax + b$, where $a \in \mathbb{R}$ and $b \in \mathbb{R}$ are trainable parameters and $\lambda \in \mathbb{R}$ is a model regularization hyper-parameter.

The loss function is designed to capture not only the personalized difference between positive and negative items but also the global differences between items through $f(S(v^+, v^-))$. Geometrically, the loss motivates the reduction of the hyperbolic distance between $u$ and $v^+$ and increases the distance between pairs $(u, v^-)$ and $(v^+, v^-)$ (see Figure \ref{fig:model}). Due to the adaptive margin $f(S(v^+, v^-))$, our model learns to better discern hyperbolic representations of positive and negative items. While \emph{HyperBPR} makes the scores for positive and negative items nearly indistinguishable (see Figure \ref{fig:comp}), \emph{TriplH} makes them significantly better separated, resulting in higher recommendation accuracy. However, we empirically observed that $f(S(v^+, v^-))$ can grow significantly, promoting the dominance of popular items and leading to trivial recommendations. To mitigate this effect, we utilize regularization $\lambda \langle v^+, v^-\rangle^2_L$.

\begin{figure}
    \centering
    \includegraphics[width=0.8\linewidth]{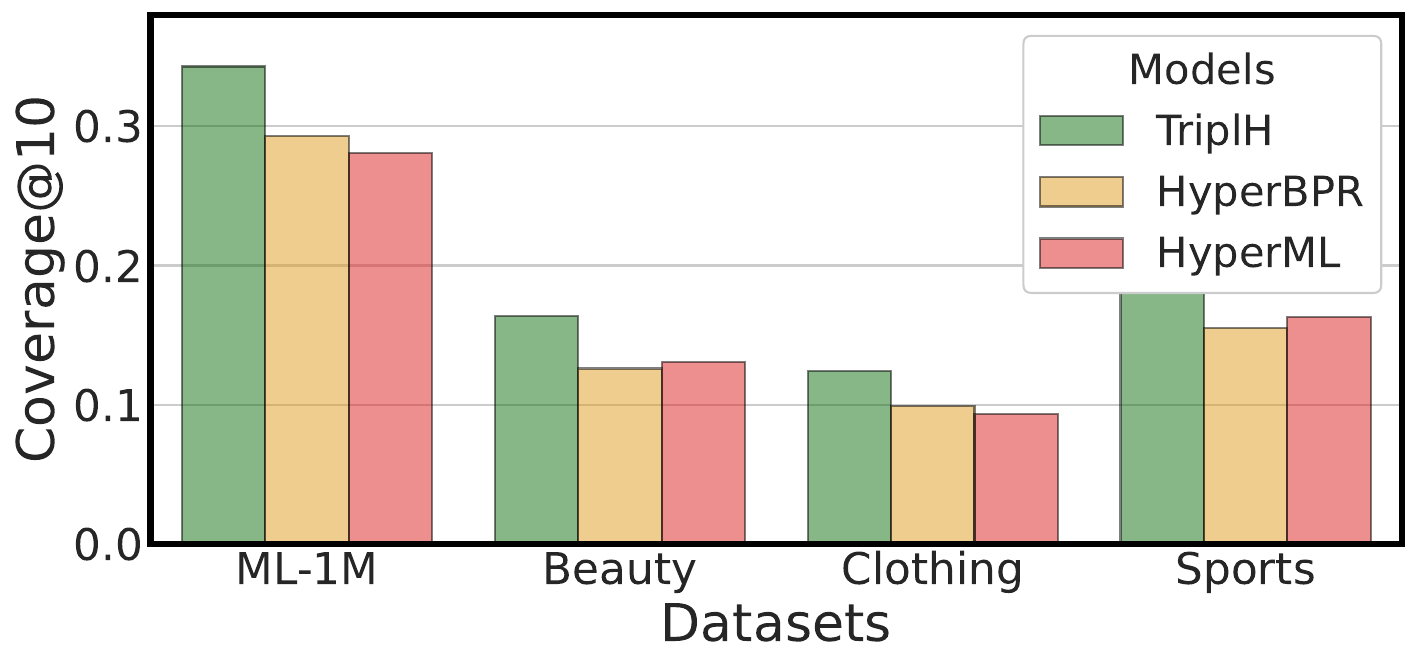}
    \vspace{-5pt}
    \caption{The best Coverage@10 of \emph{TriplH} in comparison to \emph{HyperBPR} and \emph{HyperML} on four datasets.}
    \label{fig:cov}
\end{figure}

For each user, the items are recommended according to the score:
\[
r_{uv} = S(u, v),
\]
where $u$ and $v$ are user and item embeddings. The higher the value of $r_{uv}$ the more relevant item $v$ for the user $u$. 

\textbf{Optimization process:} The model is learned with the AdamW optimizer \cite{adam}  in the Euclidean $\mathbb{R}^d$ space, and then the learned vectors are mapped to the hyperboloid via $x^\top \rightarrow [\sqrt{\beta + \|x\|_2^2}, x^\top]$. This approach offers \emph{greater computational stability} compared to learning vector representations in the Poincaré ball \cite{peng2021hyperbolic}, as it avoids computationally complex and unstable hyperbolic trigonometric functions \cite{nickel2017poincare}.

To demonstrate the effectiveness of learning user and item representations in hyperbolic space over Euclidean space, we additionally learn similar to \eqref{loss} model in Euclidean space with 
\begin{equation}
S(x, y) = x^\top y, \text{ } x \in \mathbb{R}^d, y \in \mathbb{R}^d.
\label{euc}
\end{equation}

\section{Experimental Setup}
\label{set}
To compare our methods with other approaches, we train and evaluate models on four established benchmarks covering different domains: MovieLens-1M \cite{movielens} and Amazon's Beauty, Clothing, and Sports \cite{amazon}, using a leave-last-out split \cite{meng2020exploring}. The dataset statistics are summarized in Table \ref{table:stats}. We employ HitRate@K and NDCG@K metrics with $K \in \{5, 10\}$ without target items subsampling.
We compare our \emph{TriplH} model \eqref{euc} with several commonly used Euclidean baselines: \emph{Matrix Factorization (MF)}, \emph{iALS} \cite{rendle2022revisiting}, \emph{SANSA} \cite{spivsak2023scalable} and \emph{BPR} \cite{rendle2012bpr}, as well as hyperbolic ones: \emph{LorentzFM} \cite{xu2020learning}, \emph{HyperBPR} \cite{chamberlain2019scalable} and \emph{HyperML} \cite{vinh2020hyperml}. To demonstrate the effectiveness of hyperbolic embeddings, we build the Euclidean \emph{TriplH} \eqref{euc} named \emph{TriplE}.
\vspace{-10pt}
\begin{table}[h]
\caption{Dataset statistics.}

	\label{table:stats}
	\footnotesize
	\centering
        \scalebox{0.95}{
	\begin{tabular}{lcccc}
		\toprule
  
		\textsf{Statistic} 
         & ML-1M &   Beauty  &  Clothing &  Sports \\
		\midrule
            \#Users & 6,040 & 22,363 & 35,598 & 39,387 \\ 
            \#Items & 3,706 & 12,101 & 18,357 & 23,033 \\
            \#Actions & 1,000,209 & 198,502 & 296,337 & 278,677 \\ 
            Avg. length & 165.60 & 8.88 & 7.08 & 8.33 \\
		\bottomrule
	\end{tabular}}

\end{table}

\begin{table*}[h]
\centering
\caption{The comparison of various approaches across four datasets is presented using HitRate (H) and NDCG (N) metrics, expressed as percentages. The best-performing model is highlighted in \textbf{bold}, while the second-best result is \underline{underlined}.}
\vspace{-5pt}
\label{tab:results}
\scalebox{0.9}{
\begin{tabular}{lcccccccccccccccc}
\toprule
 & \multicolumn{4}{c}{ML-1M} & \multicolumn{4}{c}{Beauty} & \multicolumn{4}{c}{Clothes} & \multicolumn{4}{c}{Sports} \\
\cmidrule(lr){2-5} \cmidrule(lr){6-9} \cmidrule(lr){10-13} \cmidrule(lr){14-17}
Model & N@5 & N@10 & H@5 & H@10 & N@5 & N@10 & H@5 & H@10 & N@5 & N@10 & H@5 & H@10 & N@5 & N@10 & H@5 & H@10 \\
\midrule
MF    & 2.21 &3.09 &3.15 &4.95 &1.04 &1.37 &1.30 &2.43 &0.43 &0.61 &0.67 &0.90 &1.02 &1.48 &1.62 &2.65 \\
BPR   &3.15 &4.41 &4.35 &6.46 &1.21 &2.13 &2.01 &3.24 &0.48 & 0.79& 0.81& 1.13& 1.11& 1.57& 1.68& 2.85\\
iALS  &2.53 &3.71 &3.65 &5.12 &1.13 &2.03 &1.88 &2.84 &0.43 &0.65 &0.67 &0.95 &1.08 &1.57 &1.61 &2.71 \\
SANSA & 3.06 & 4.17 & 4.21 & 7.01 & 1.21 & 2.16 & 2.12 & 3.25 & 0.47 & 0.76 & 0.80 & 1.14 & 1.07 & 1.68 & 1.74 & 2.93 \\  
LorentzFM  &2.68 &3.83 &4.01 &5.56 &1.15 &2.05 &1.88 &3.08 &0.45 & 0.71& 0.75& 1.06& 1.04& 1.48& 1.64& 2.81\\
HyperML  &\underline{3.27} &\underline{4.76} &\underline{4.91} &\underline{7.14} &1.32 &2.38 &2.32 &3.43 &0.53 & 0.84& 0.89& 1.23& \underline{1.17}& \underline{1.81}& \underline{1.85}& \underline{3.03}\\
HyperBPR  &3.14 &4.61 &4.63 &7.06 &\underline{1.47} &\underline{2.44} &\underline{2.41} &\underline{3.75} &\underline{0.54} & \underline{0.87}& \underline{0.92}& \underline{1.27}& 1.13& 1.75& 1.76& 2.98\\
\midrule
TriplE  &3.21 &4.45 &4.54 &6.57 &1.17 &2.13 &1.96 &3.20 &0.48 & 0.81& 0.82& 1.15& 1.12& 1.65& 1.71& 2.87\\
TriplH  &\textbf{3.39} &\textbf{4.92} &\textbf{5.07} &\textbf{7.56} &\textbf{1.60} &\textbf{2.51} &\textbf{2.49} &\textbf{3.89} &\textbf{0.57} & \textbf{0.92}& \textbf{0.95}& \textbf{1.34}& \textbf{1.27}& \textbf{1.96}& \textbf{1.98}& \textbf{3.24}\\ 
\midrule
Improv. \% & 3.7 & 3.4 & 2.3 & 5.9 & 8.8 & 2.9 & 3.3 & 3.7 & 5.6 & 5.7 & 3.3 & 5.5 & 8.5 &  8.3 & 7.0 & 6.9\\
\bottomrule
\end{tabular}
}
\end{table*}

\begin{figure}
    \centering
    \includegraphics[width=0.8\linewidth]{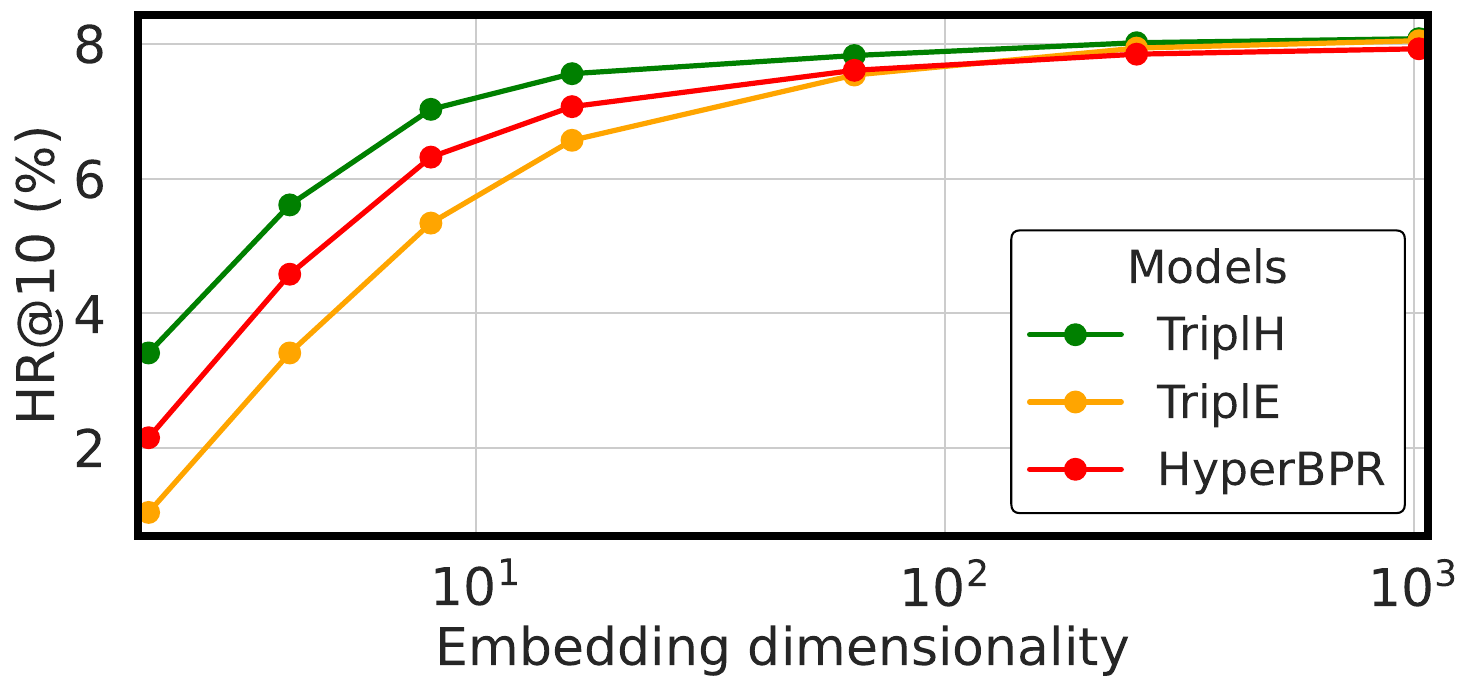}
    \vspace{-7pt}
    \caption{The improved performance (HR@10 in \%) of \emph{TriplH} in comparison to \emph{TriplE} and \emph{HyperBPR} models on the MovieLens-1M dataset for different embedding sizes $d$.}
    \label{fig:hvse}
\end{figure}

\section{Results}
\label{res}
As shown in Table \ref{tab:results}, which presents the model comparison results, our approach outperforms other Euclidean and hyperbolic methods across all metrics and datasets, achieving improvements of up to $\textbf{7\%}$ on some of them (the results obtained are statistically significant). Incorporating additional item-item interaction term $f(S(v^+, v^-))$ significantly enhances the quality of recommendations. Furthermore, the hyperbolic \emph{TripleH} model surpasses the Euclidean \emph{TriplE}, demonstrating a better expressiveness of the hyperbolic vector representations. Moreover, our adaptive hyperbolic margin enhances recommendation quality compared to the \emph{HyperML} \cite{vinh2020hyperml} and the \emph{HyperBPR} \cite{chamberlain2019scalable} approaches, as it captures additional dependencies between positive and negative items for each user, leading to a more accurate learning of user preferences.

We evaluate the diversity of recommendations of the hyperbolic methods, to compare the Coverage@10 metric, which measures the proportion of recommended items to the total catalogue size, for the \emph{HyperBPR}, \emph{HyperML}, and \emph{TriplH} approaches. As shown in the Figure \ref{fig:cov}, the \emph{TriplH} significantly outperforms the other models in recommendations diversity. Moreover, we observed that the share of medium popular items \cite{steck2011item} among all recommendations is the highest among other hyperbolic approaches, which means that the model reduces the effect of popularity bias \cite{klimashevskaia2024survey}. It could be explained by the fact that item-item relation term $f(S(v^+, v^-))$ with regularization $\lambda \langle v^+, v^-\rangle^2_L$ \eqref{loss} learns the preferences of users and reduces the domination of popular items in the recommendations. 

To demonstrate the superior effectiveness of hyperbolic embeddings compared to Euclidean ones, we compare the performance of our hyperbolic \emph{TriplH} model with the Euclidean \emph{TriplE} baseline across different embedding dimensions $d$. As shown in Figure~\ref{fig:hvse}, the Lorentzian model achieves significantly better performance than the Euclidean one, particularly for small embedding sizes. This experiment demonstrates greater representational capacity of hyperbolic spaces, confirming the conclusions of prior research~\cite{peng2021hyperbolic,kruk}. Consequently, our model remains highly effective even under strict constraints on embedding dimensionality, total parameter count, or when applied to data with low latent dimensionality, making it particularly suitable for resource-constrained applications.

\begin{figure}
    \centering
    \includegraphics[width=0.8\linewidth]{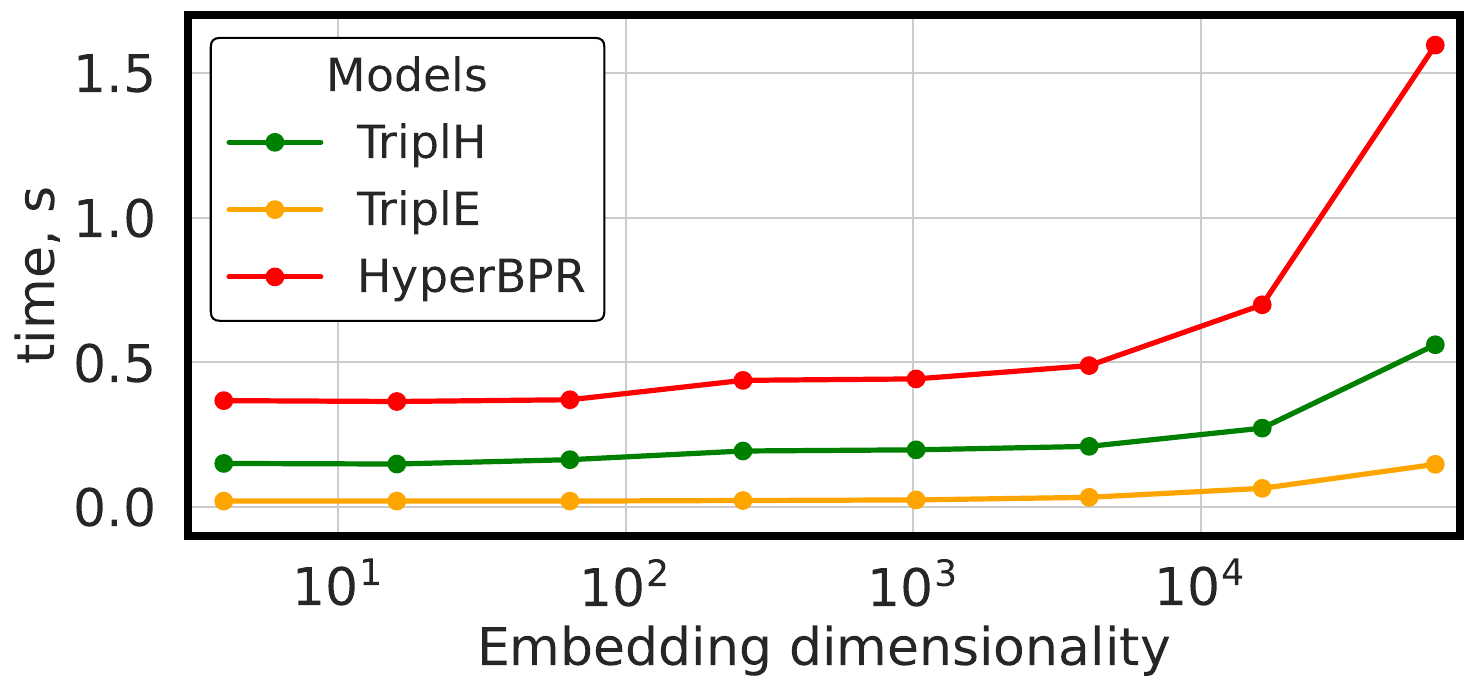}
    \vspace{-7pt}
    \caption{The improved inference time of \emph{TriplH} in comparison to \emph{TripleE} and \emph{HyperBPR} models.}
    \label{fig:lvsp}
\end{figure}
We also compare the inference times of \emph{HyperBPR} and \emph{TripleH} to demonstrate the efficiency of the hyperbolic embeddings derived from the Lorentzian model as opposed to those from the Poincaré ball. As shown in Figure \ref{fig:lvsp}, the inference time of the \emph{TripleH} model using the Squared Lorentz distance \eqref{square} is up to $\textbf{3}$ times lower than that of \emph{HyperBPR} or \emph{HyperML} with the Poincaré distance, as \emph{HyperBPR} and \emph{HyperML} have an absolutely similar inference procedure and comparable inference times. This difference can be explained by the fact that the Poincaré models involve exponential mapping and computationally intensive operations with hyperbolic trigonometric functions \cite{peng2021hyperbolic}. In contrast, operations associated with the Squared Lorentz distance are significantly faster. Additionally, the operations in the Squared Lorentz distance are more numerically stable, which contributes to the higher reliability and robustness of our \emph{TriplH} approach.

\vspace{-2pt}
\section{Conclusion}
\label{con}
In this work, we introduce \emph{TriplH}, a geometrically inspired hyperbolic approach for modelling ternary item-user-item interactions in recommender systems. This approach yields more expressive hyperbolic representations of users and items compared to Euclidean and other hyperbolic embeddings, leading to enhanced recommendation performance over the other Euclidean and hyperbolic approaches. Moreover, our model achieves this with lower-dimensional embeddings than those required in Euclidean space. The proposed item-item interaction term effectively captures dependencies between items for each user, helping to reduce popularity bias and promote diversity in recommendations. Additionally, leveraging Lorentzian embeddings, we achieve $3$ times faster inference times and greater computational stability relative to Poincaré ball-based methods, making our approach both more reliable and efficient.

\begin{acks}
Support from the Basic Research Program of HSE University is gratefully acknowledged. The calculations
were performed in part through the computational resources of HPC facilities at HSE University \cite{kostenetskiy2021hpc}.
\end{acks}

\bibliographystyle{ACM-Reference-Format}
\balance
\bibliography{submission_TriplH}

\end{document}